\include{graphicsx} 

\documentclass[12pt,preprint]{aastex}

\slugcomment{accepted for publication in the Astrophysical Journal on Feb 20, 2012}

\shorttitle{LBT AO observations of the Trapezium}
\shortauthors{Close et al.}

\begin{document}


\title{High Resolution Images of Orbital Motion in the Orion Trapezium Cluster with the LBT AO System{\footnotemark}}

\footnotetext{$^1$The LBT is an international collaboration among institutions in the United States, Italy and Germany. LBT Corporation partners are: The University of Arizona on behalf of the Arizona university system; Istituto Nazionale di Astrofisica, Italy; LBT Beteiligungsgesellschaft, Germany, representing the Max-Planck Society, the Astrophysical Institute Potsdam, and Heidelberg University; The Ohio State University, and The Research Corporation, on behalf of The University of Notre Dame, University of Minnesota and University of Virginia.}


\author{Close, L.M.$^1$, Puglisi, A$^{2}$, Males, J.R.$^1$, Arcidiacono, C$^6$,  Skemer, A$^1$, Guerra, J.C.$^3$, Busoni, L.$^2$, Brusa, G.$^{3}$, Pinna, E.$^2$, Miller, D.L.$^3$, Riccardi, A.$^2$, McCarthy, D.W.$^1$, Xompero, M.$^1$, Kulesa, C.$^1$, Quiros-Pacheco, F.$^2$, Argomedo, J.$^2$, Brynnel, J.$^3$, Esposito, S.$^2$, Mannucci, F.$^2$, Boutsia, K.$^{3,4}$, Fini, L.$^2$, Thompson, D.J.$^3$, Hill, J.M.$^3$, Woodward, C.E.$^5$, Briguglio, R.$^2$, Rodigas, T.J.$^1$, Briguglio, R.$^2$, Stefanini, P$^2$, Agapito, G$^2$, Hinz, P.$^1$, Follette, K.$^1$, Green, R.$^3$
}

\email{lclose@as.arizona.edu}

\affil{$^1$Steward Observatory, University of Arizona, Tucson, AZ 85721, USA}
\affil{$^2$INAF - Osservatorio Astrofisico di Arcetri, I-50125, Firenze, Italy}
\affil{$^3$LBT Observatory, University of Arizona, Tucson, AZ 85721, USA}
\affil{$^4$INAF-Osservatorio Astronomico di Roma, Via Frascati 33, I-00040, Monteporzio, Italy}
\affil{$^5$University of Minnesota Minneapolis, MN 55455 USA}
\affil{$^6$INAF - Osservatorio Astronomico di Bologna, I-40127 Bologna Italy}
%\affil{$^2$Institute for Astronomy, University of Hawaii, Honolulu, HI}
%\affil{$^3$European Southern Observatory, Garching, Germany}   



\begin{abstract} 

  The new 8.4m LBT adaptive secondary AO system, with its novel pyramid wavefront sensor, was used to produce very high Strehl ($\ga75\%$ at
  2.16$\mu m$) near infrared narrowband ($Br\gamma$: 2.16$\mu m$ and [FeII]:
  1.64$\mu m$) images of 47 young ($\sim 1$ Myr) Orion Trapezium
  $\theta^{1}$ Ori cluster members. The inner
  $\sim41 \times 53\arcsec$ of the cluster was imaged at spatial resolutions
  of $\sim0.050\arcsec$ (at 1.64$\mu m$). A combination of high spatial resolution and high S/N yielded relative binary positions to $\sim 0.5 $ mas
  accuracies. Including previous speckle data, we analyze a 15 year baseline of high-resolution
  observations of this cluster. We are now
  sensitive to relative proper motions of just $\sim0.3$ mas/yr
  (0.6 km/s at 450 pc) this is a $\sim7\times$ improvement in
  orbital velocity accuracy compared to previous efforts. We now detect clear orbital motions in the $\theta^{1}$ Ori $B_{2}B_{3}$ system of $4.9\pm0.3$
  km/s and $7.2\pm0.8$ km/s in the
  $\theta^{1}$ Ori $A_{1}A_{2}$ system (with correlations of PA vs. time at $>99\%$ confidence). All five members of the
  $\theta^{1}$ Ori $B$ system appear likely as a gravitationally bound ``mini-cluster''. The
  very lowest mass member of the $\theta^{1}$ Ori $B$ system ($B_{4}$; mass $\sim 0.2
  M_{\sun}$) has, for the first time, a clearly detected motion (at
  $4.3\pm2.0$ km/s; correlation=99.7\%) w.r.t $B_1$. However, $B_4$ is most
  likely in an long-term unstable (non-hierarchical) orbit and
  may ``soon'' be ejected from this ``mini-cluster''. This ``ejection'' process could play a major role in the
  formation of low mass stars and brown dwarfs.

\end{abstract}


\section {INTRODUCTION}

The detailed formation of stars is still a poorly understood process. In
particular, the formation mechanism of the lowest mass stars and brown dwarfs is
uncertain. Detailed 3D (and N-body) simulations of star formation by \cite{bat02,bat03,bat09,bat11} and \cite{par11} all
suggest that stellar embryos frequently form into ``mini-clusters'' which
dynamically decay, ``ejecting'' the lowest mass members. Such theories
can explain why there are far more field brown dwarfs (BD) compared to
BD companions of solar type stars \citep{mcc03} or early M stars
\citep{hin02}. Moreover, these theories which invoke some sort of
dynamical decay \citep{dur01} or ejection \citep{rep02} suggest that
there should be no wide ($>20$ AU) very low mass (VLM; $M_{tot}<0.185
M_{\sun}$) binary systems observed in the field (age $\sim 5$ Gyr). Indeed, the AO surveys of
\cite{clo03a} and the HST surveys of \cite{rei01a,bur03,bou03,giz03} have not
discovered more then a few wide ($>16$ AU) VLM systems of the systems in the field population (for a review see \cite{bur07}). Additionally, the dynamical biasing towards the ejection of the
lowest mass members naturally suggests that the frequency of field VLM
binaries should be much lower ($\la 5\%$ for $M_{tot}\sim 0.16
M_{\sun}$) than for more massive binaries ($\sim 60\%$ for $M_{tot}
\sim 1 M_{\sun}$). Indeed, observations suggest that the binarity of
VLM systems with $M_{tot} \la 0.185 M_{\sun}$ is $10-15\%$
\citep{clo03a, bur03, bur07} which, although higher than predicted is still
lower than that of the $\sim 60\%$ of G star binaries \cite{duq91}. However has is noted in \cite{clo07} there is evidence that in young clusters wide VLM binaries are much more common than in the old field population. They attribute this to observing these wide VLM systems before they are destroyed by encounters in their natal clusters. Hence, we need to look at nearby young clusters to see these low-mass objects before ejection has occurred.

Despite the success of these decay or ejection scenarios in predicting
the observed properties of low mass VLM stars and binaries , it is still not clear that
``mini-clusters'' even exist in the early stages of star formation. To
better understand whether such ``mini-clusters'' do exist we have
examined the closest major OB star formation cluster for signs of such
``mini-clusters''. Here we focus on the $\theta^1$ Ori stars in the
famous Orion Trapezium cluster. Trying to determine if some of the tight star
groups in the Trapezium cluster are gravitationally bound is a first
step to determining if bound ``mini-clusters'' exist. Also it is
important to understand the true number of real, physical, binaries in
this cluster, as there is evidence that the overall number of binaries
is lower (at least for the lower mass members) in the dense trapezium
cluster compared to the lower density young associations like
Taurus-Auriga \citep{mcc11,koh06}.  In particular, we will examine the
case of the $\theta ^1$ Ori A and B groups in detail.

The Trapezium OB stars ($\theta^1$ Ori A, B, C, D, and E) consists of
the most massive OB stars located at the center of the Orion Nebula
star formation cluster (for a review see \cite{gen89}). Due to the
nearby and luminous nature of these stars they have been the target of several
high-resolution imaging studies. Utilizing only tip-tilt compensation
\cite{mcc94} mapped the region at $K^\prime$ from the 3.5-m Calar Alto
telescope. They noted that $\theta^1$ Ori B was really composed of 2
components ($B_1$ \& $B_2$) about $\sim 1\arcsec$ apart. Higher $\sim
0.15\arcsec$ resolutions were obtained from the same telescope by
\cite{pet98} with speckle holographic observations. At these higher
resolutions \cite{pet98} discovered that $\theta^1$ Ori $B_2$ was
really itself a $0.1\arcsec$ system ($B_2$ \& $B_3$) and that
$\theta^1$ Ori A was really a $\sim0.2\arcsec$ binary ($A_1$ \&
$A_2$). A large AO survey of the inner 6 square arcminutes was carried
out by \cite{sim99}, who discovered a very faint (100 times fainter
than $B_1$) object ($B_4$) located just $0.6\arcsec$ between $B_1$ and
$B_2$. Moreover, a spectroscopic survey \citep{abt91} showed that
$B_1$ was really an eclipsing spectroscopic binary ($B_1$ \& $B_5$;
sep. 0.13 AU; period 6.47 days). As well, $\theta^1$ Ori $A_1$ was
also found to be a spectroscopic binary ($A_1$ \& $A_3$; sep. 1 AU;
\cite{bos89} ). \cite{wei99} carried out bispectrum speckle
interferometric observations at the larger Russian SAO 6-m telescope
(2 runs in 1997 and 1998). These observations showed $\theta^1$ Ori C
was a very tight 0.033$\arcsec$ binary. These observations also
provided the first set of accurate relative positions for these
stars. \cite{sch03} has continued to monitor this cluster of stars and
detected an orbital motion (of $\Delta PA \sim
6^{\circ}$ for $\theta^{1}$ Ori $A_2$ around $A_1$ and a $\Delta PA$
of $\sim 8^{\circ}$ for $\theta^{1}$ Ori $B_3$ around $B_2$ over a 5.5 yr
baseline). 

\cite{clo03c} utilized the Gemini telescope (with the
Hokupa'a AO system) and then observed $\theta^1$ Ori B during
commissioning of the first adaptive secondary deformable mirror at the
6.5-m MMT telescope. This extended the baseline by 2 years. Now during the science verification of the world's first ``next generation'' adaptive secondary mirror with the LBT AO System the full Trapezium cluster was observed again with excellent performance (K Strehl $\ga 75\%$). Now we have over 14 years of observations of this field with at $<0.08\arcsec$ resolutions.

In this paper we outline how these LBT observations
were carried out with the relatively new PISCES camera and LBT AO system (FLAO). We detail how these data were calibrated and reduced and how the stellar positions were measured. We fit
the observed positions to calculate velocities (or upper limits) for
the $\theta^1$ Ori B \& A stars. 
While \cite{sch03} and \cite{clo03c} had hints that the $\theta^1$ Ori B 
group may be a bound ``mini-cluster'' ---we show it is clearly so, with the first detection of orbital motion of the lowest mass member.

\section{INSTRUMENTAL SET-UP}

 We utilized the LBT adaptive secondary AO
 system to obtain the most recent highest-resolution (unsaturated) images of the young
 stars in the Trapezium cluster (the $\theta^{1}$ Ori group). This is not a
simple task, since as telescopes have increased in size and (with AO) Strehl, so now the bright stars tend to saturate --in even the shortest exposures. Hence, special precautions are needed to avoid saturation of the bright Trapezium stars themselves. It is now almost impossible to make unsaturated, but diffraction-limited, images of the bright Trapezium stars with modern 8m class AO systems at high Strehl. Witness the fact that this is the first such dataset published in 9 years. Hence this dataset is unusually important. The next subsections outline how this was accomplished.

\subsection{The LBT AO System}

The 8.4m LBT telescope has a unique ``first light adaptive optics'' (FLAO) system. To reduce
the aberrations caused by atmospheric turbulence all AO systems have a
deformable mirror which is updated in shape at $\sim 500$ Hz. Except for the MMTAO system \citep{wil03,hin10},
all adaptive optics systems have located this deformable mirror
(DM) at a re-imaged pupil (effectively a compressed image of the
primary mirror). To reimage the pupil onto a DM typically requires 6-8
warm additional optical surfaces, which significantly increases the
thermal background and decreases the optical throughput of the system
\citep{llo00}. However, the LBT utilizes a next generation adaptive secondary DM.
%(see Figure \ref{fig0})
 This DM is both the secondary mirror of the
telescope and the DM of the AO system (like with MMTAO). 
%(see Figure \ref{fig1}).
 In this
manner there are no additional optics required in front of the science
camera. Hence the emissivity is lower. The LBT's DM is a much more advanced ``second generation'' adaptive secondary mirror (ASM), which enables the highest on-sky Strehl ($>80\%$ at H band) of any 8-10m telescope today \citep{esp11}. 

The LBT ASM consists of 672 voice coil actuators that push (or pull) on 672 small
magnets glued to the backsurface of a thin (1.6 mm), 0.911 m
aspheric ellipsoidal Zerodur glass ``shell'' (for a detailed review of the secondary mirror see \cite{esp10,esp11}).
%(see Figure \ref{fig0}).%
 We have complete
positional control of the surface of this reflective shell by use of
a 70kHz capacitive sensor feedback loop. This positional feedback loop
allows one to position an actuator of the shell to within $\sim5$ nm rms (total wavefront surface 
errors amount to only $\sim50$ nm rms over the whole secondary). The AO
system samples (and drives the ASM) at 990 Hz using 400 active controlled modes (with 672 actuators) on bright stars (R<8 mag). 

The wavefront slopes are measured with the very accurate (and well
calibrated, with low aliasing error) Pyramid Wavefront sensor
(PWFS). This is the first large telescope to use a PWFS. The performance
of the FLAO PWFS is excellent. The uniquely low residual wavefront
errors obtained by the PWFS + ASM combination is due, in part, to the
very accurate (high S/N) interaction matrix that can be obtained in closed-loop
daytime calibrations with a retro-reflecting optic that takes advantage of the
Gregorian (concave) nature of the secondary. To guarantee strict ``on-sky''
compliance with the ``daytime calibrated'' interaction matrix pupil/ASM/PWFS geometry the PWFS utilizes a novel
``closed-loop pupil alignment system'' that maintains the pupil
alignment to $<2.5\mu m$ during all closed-loop operations on bright stars (like our Trapezium guide stars). For a
detailed review of the LBT FLAO AO system see
\cite{esp11} and references within. 

\subsection{The LBT AO Observations}

During LBT science verification (the last AO commissioning run of the first LBT ASM) we observed the $\theta^{1}$ Ori B and C fields on the night of Oct 16, 2011 (UT). The AO system corrected the
lowest 400 system modes and was updated at 990 Hz. Without AO correction our
images had FWHM=$\sim0.5\arcsec$ at [FeII], after AO correction our final 10 min image
achieved FWHM=$0.050\arcsec$ (close to the diffraction limit of
$0.041\arcsec$ at [FeII] (1.644 $\mu m$)). 

%A detailed analysis suggested that
%during our engineering run a 40 Hz vibration in the MMT telescope
%increased our FWHM by $\sim 0.015\arcsec$ and decreased our Strehl by
%a factor of two. We are in the process of identifying and decreasing
%the effect of this 40 Hz vibration. In any case, as Figure \ref{fig2}
%clearly shows, there is a large improvement in image quality (the
%Strehl increases by 20 times) with the adaptive secondary AO system.

\subsection{ The PISCES NIR Camera}

These observations utilized the first light AO science camera, PISCES, which has been modified for the LBT AO system.  PISCES has a 1024x1024 1-2.5 $\mu m$ HAWAII array. Here we
used the narrowband $Br\gamma$ (2.16 $\mu m$) and [FeII] (1.644 $\mu m$)
filters to minimize saturation on the array. We also utilized a warm
25 mm dia. neutral density filter (ND2: with 1\% transmission) which
was custom mounted by flexible adhesive strips on the PISCES dewar within a few mm of the f/15
focal plane of the ASM. Since the high quality flat ND2 was nearly in the warm
first focal plane (in front of the PISCES dewar window) it cannot
significantly alter, or distort, the platescale or optical quality of PISCES.

The PISCES focal plane platescales were calibrated by the astrometry of
seven single (relatively faint) stars
\footnote{Typically the stars in the Trapezium used for this platescale test move at only $\sim0.0015\arcsec$/yr so the platescale error over a $5\arcsec$ distance is $\sim2x10^{-5}\%$ error --- which is much smaller than the rms fitting platescale errors of $\sim0.25\%$.} 
from $3.4-8.5\arcsec$ from $\theta^1$ Ori C (see sections 3 and 4 for more details about how the images were first distortion corrected and combined etc.). We note here that we only used the data where $\theta^1$ Ori C was the guide star for the platescale calibration. The 3 dither frames that used this guide star were all taken within 20 min of each other, and so the PA angle is assumed to be fixed for all exposures.

The positions (found by IRAF {\it allstar} PSF fitting) of these seven stars from our LBT AO images were compared to unsaturated
HST ACS WCS astrometry from the publicly available archived data of \cite{ric07}.  Platescales and rms errors were then determined for the $Br\gamma$
and [FeII] filters with the IRAF {\it geomap} task. The {\it geomap} task found $0.019350\pm0.000047\arcsec$/pix $Br\gamma$ platescale
(providing a $19.8\times 19.8\arcsec$ FOV). At
[FeII] the platescale was slightly finer at
$0.019274\pm0.000036\arcsec$/pix. For PISCES it is well known that, due to a slight focus change, the bluer wavelengths have a slightly finer platescale.

The steps used to align the Y axis of the PISCES images (which were all taken with
the rotator following) it was first necessary to flip the image about the X
axis. Then each image was rotated (with the IRAF {\it rotate} task) by the fixed PA+90 of the rotator
(POSANGLE FITS keyword value +90 degrees). All 3 rotated images were then combined. At this point it was found by {\it geomap} in the
$Br\gamma$ image that the direction of North was slightly ($0.898^{\circ}$)
East of PISCES's Y axis compared to the HST image. Hence a final
rotation of $-0.898^{\circ}$ was applied to the final image. At [FeII]
this additional rotation was a very similar $-0.959^{\circ}$ value. The rms uncertainty adopted for the LBT rotator angle is estimated as $\sim 0.3^{\circ}$ which dominates the small $0.062^{\circ}$ error between the two geomap solutions. We conclude that most of the $\sim 0.3^{\circ}$ is due to systematic errors in the distortion corrections. 

The camera was mounted under a high optical quality dichroic
which sent the visible light (0.5-1 $\mu m$) to the 30x30 subaperture 
Pyramid wavefront sensor (PWFS; \cite{esp10a}). The PWFS communicates with the the ASM mirror (which has 672 actuators, but $\sim 20$ where not operational on this run, with no real loss in performance). The infrared light ($\lambda >
1 \mu m$) was transmitted through the dichroic to PISCES.

\section{OBSERVATIONS \& REDUCTIONS}

For the $\theta^{1}$ Ori C field we locked the AO system (at 990Hz, 400
modes) on the bright O5pv star $\theta^{1}$ Ori C (V=5.13 mag) and
dithered over 3 positions on the PISCES array with a short set of
10x0.8 second unsaturated exposures (save C1 which at H=4.48 saturated
even in [FeII] with the ND2). Immediately following the unsaturated
exposures a set of 10x20 second exposures were obtained at each dither
position. This whole procedure was repeated with in $Br\gamma$ with slightly
different dithers. Hence for a few sources on the edges of the frames
there is only photometry in one filter. However, when reduced both
filter images covered an area of $\sim31\times24\arcsec$ centered on
$\theta^{1}$ Ori C.  We note that $\theta^{1}$ Ori C is really a $\sim0.04"$ binary composed of C1 and C2, (see \cite{kra07} for more details). Due to the very red nature of the Trapezium
sources we only needed half the number of images at $Br\gamma$ compared to
[FeII]. Hence we obtained 3x5x0.8 second unsaturated exposures and
3x5x20 second deeper exposures at $Br\gamma$.

Then the AO system was locked on the nearby star $\theta^{1}$ Ori B1
(V=7.96 mag) and was dithered over a similar sized area to produce a
set of 3x10x0.8 s unsaturated images [FeII] images followed by
3x10x20 s deeper images at [FeII]. Again half that data was obtained at
$Br\gamma$ (3x5x0.8 s and 3x5x20 s). The final image in the $\theta^{1}$ Ori B
field spanned $\sim31\times24\arcsec$ centered on $\theta^{1}$ Ori B1.

Before the images were combined each image was first corrected with
PISCES appropriate values for the {\it corquad} ``overshoot'' program
\footnote{{\it
  http://aries.as.arizona.edu/$\sim$observer/dot.corquad.pisces}}. Then
they were distortion ``pincushion'' corrected with the cubic distortion
solution for PISCES from a pinhole array mask\footnote{{\it
  http://wiki.lbto.arizona.edu/twiki/pub/AdaptiveOptics/PiscesDistorsion/pisces.cubic}}
and the IRAF {\it drizzle} task. At this point the individual frames
could be reduced in a normal manner. We used our custom AO image
reduction script of \cite{clo03a} to flat field, sky subtract,
cross-correlate, and median combine each image. The final deep images
of the C and B fields had a total exposure time of 10 min in the
overlap region in the [FeII] images and 5 min in the $Br\gamma$ images. The unsaturated images had total exposure times of 24 s and 12 s in [FeII] and $Br\gamma$ respectively. Each of these reduced images were then flipped in X, rotated by POSANGLE+90 degrees, and corrected by -0.9 degrees --as described above to match the HST ACS astrometry with {\it geomap} to that of \cite{ric07}. Hence North is up and East is to the left in each of the final images.

\begin{figure}
\includegraphics[angle=0,width=\columnwidth]{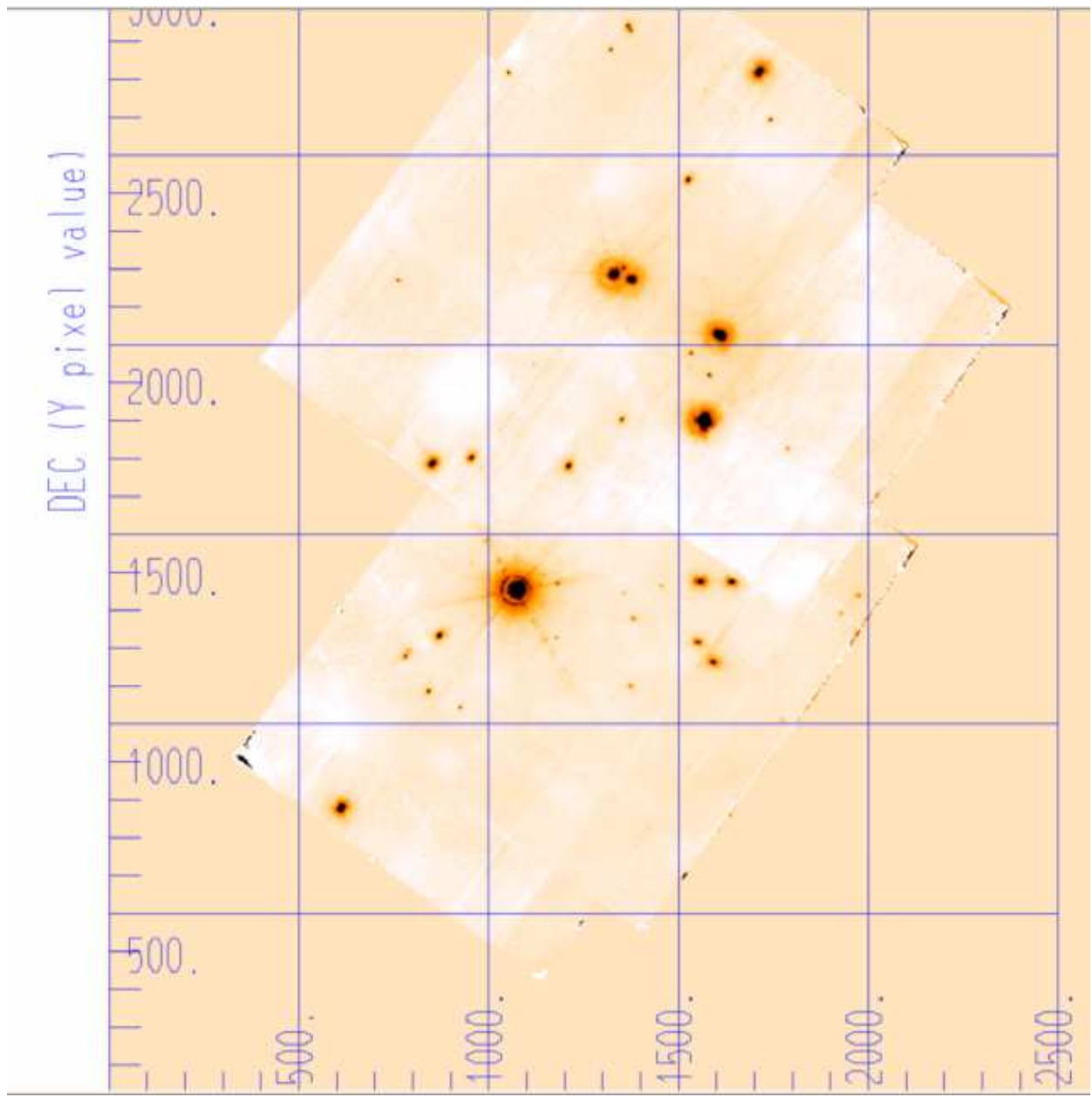}\caption{
The  $\theta^{1}$ Ori cluster as imaged over $\sim 41\times 53\arcsec$ FOV at LBT with the LBT AO and PISCES in $Br\gamma$. Logarithmic color scale. North is up and east is left. Note that the ``rings'' around the stars are limit of the ASM's control radius $r_c = \lambda/2d\sim0.8\arcsec$ (where the space between AO actuators maps to roughly d$\sim0.27$m on the LBT primary). Inside this control radius the FLAO system can ``carve out'' the PSF to reveal faint companions. This is the first image ever obtained which shows a field of multiple stars each with a ``dark ring'' in its PSF. Each pixel is 19.35 mas, see Table 1 for a complete list of the photometry and astrometry for this field. North is up, East to the left, and the X Y grid is in pixels --corresponding to the pixel values listed in Table 1. Hence this figure's XY grid can be used to located any object from its (X,Y) coordinates in Table \ref{tbl-0}}
\label{fig1}
\end{figure}

\begin{figure}
\includegraphics[angle=0,width=\columnwidth]{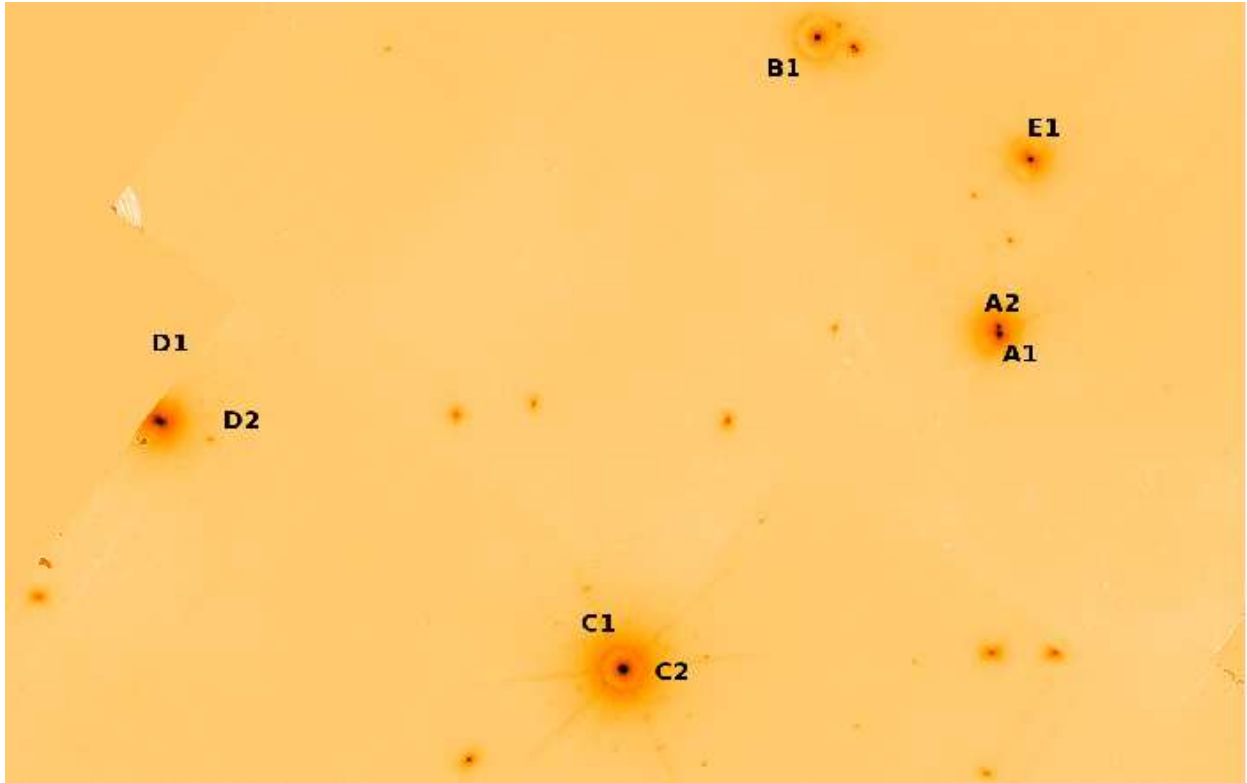}\caption{
The locations and nomenclature of \cite{clo03c} of the $\theta^{1}$ Ori Trapezium stars as imaged over $\sim35\times 30\arcsec$ FOV at LBT with LBTAO/PISCES in [FeII]. Logarithmic color scale. North is up and east is left. Note that the object ``$A_1$'' is really a spectroscopic binary ($A_1A_3$); where the unseen companion $A_3$ is separated from $A_1$ by 1 AU \citep{bos89}. The B group is shown in more detail in Figs. \ref{fig4} - \ref{lbt_B}. It is not currently clear if D2 is physically related to D1. E1 appears to be a single star. No new faint companions were discovered (at $>5\sigma$) around any of the Trapezium stars down to brown dwarf masses ($Br\gamma<16.0$ which converts to $K<14$ mag) at separations $\ga 0.1\arcsec$.
}
\label{fig2}
\end{figure}

While all astrometry and photometry was performed on these individual fields (on unsaturated images), the last step was combining these two C and B fields into a single large image of the entire Trapezium. The final images (see Figs. \ref{fig1} and \ref{fig2}) have a size of $\sim41\times53\arcsec$. This image is the largest AO image obtained by the LBT to date.

\section{ASTROMETRY \& PHOTOMETRY}

All eight reduced images (the C and B fields at both [FeII] and $Br\gamma$ at
both long and short exposures) were analyzed with the
DAOPHOT PSF fitting task {\it allstar} \citep{ste87}. The
photometry and astrometry are summarized in Table \ref{tbl-0}. The
columns of Table \ref{tbl-0} are self explanatory. We note that the
zeropoints for the [FeII] and $Br\gamma$ photometry were arbitrary as these are
narrow band filters. However, it is clear that the $\Delta$ magnitudes
at [FeII] seem to closely track the true $\Delta H$ magnitudes, since
these sources are mainly continuum at $1.64 \mu m$. However, these sources are all accreting and have excess emission at $Br\gamma$. Hence, the [FeII]-$Br\gamma$ color given in the last column of Table \ref{tbl-0} is largest for objects that have very active accretion and/or a high level of circumstellar absorption.

\begin{figure}
\includegraphics[angle=0,width=\columnwidth]{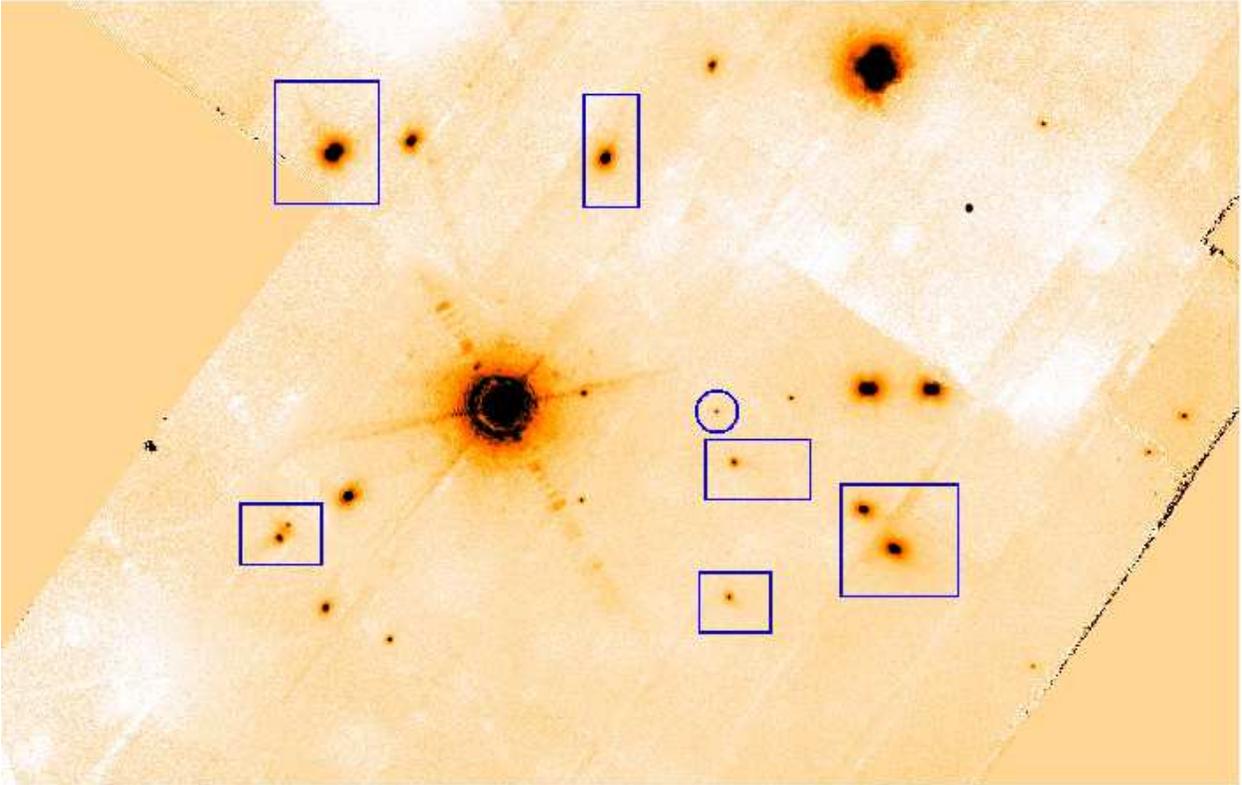}\caption{
The center of $\theta^{1}$ Ori C as imaged over $25\times 20\arcsec$ FOV at LBT with LBTAO/PISCES in $Br\gamma$. Logarithmic color scale. North is up and east is left. Note the blue squares have $Br\gamma$ ``tails'' away from C, and the circle is around a very red $Br\gamma$ object that is much brighter at $Br\gamma$ than [FeII] or H band. 
}
\label{fig3}
\end{figure}

Note in the the deep $Br\gamma$ image, well known ``tails'' of emission point away from $\theta^{1}$ Ori C, and are highlighted by blue rectangles in Fig. \ref{fig3} and in the comments column of Table \ref{tbl-0}. Also each tight binary is called out in the table as well as each member of the bright Trapezium stars themselves. All the tight ($<0.5\arcsec$) binaries have additional details in Table \ref{tbl-2}.

We minimized the small anisoplanatic PSF radial fitting errors in table \ref{tbl-0} by using a spatially variable PSF in the DAOPHOT {\it psf} task. The most heavily weighted  
PSF star used was unsaturated $\theta^{1}$ Ori $B_{1}$ itself. Since all the members of
the $\theta^{1}$ Ori $B$ group are located within $1\arcsec$ of
$\theta^{1}$ Ori $B_{1}$ the PSF fit is particularly excellent there (there is no
detectable change in PSF morphology due to anisoplanatic effects
inside the $\theta^1$ Ori B group \citep{dio00}). Moreover, the residuals over the whole field were less than a few \% after PSF subtraction. This is not really surprising given the quality of the night combined with the fact that no star was further than $\sim8\arcsec$ from the guide star. However, to minimize this affect, we only used the longer wavelength $Br\gamma$ astrometry in Table \ref{tbl-0} where anisoplanatic PSF effects were much less significant. 

The relative positional accuracy is an excellent $\sim 0.5 $ mas for the bright binaries that are tighter than $0.2\arcsec$, but the absolute RA and DEC positions given in Table \ref{tbl-0} are typically only good to $\la0.1\arcsec$ due to the $0.25\%$ uncertainty in the platescale for this new instrument.

\begin{figure}
\includegraphics[angle=0,width=0.4\columnwidth]{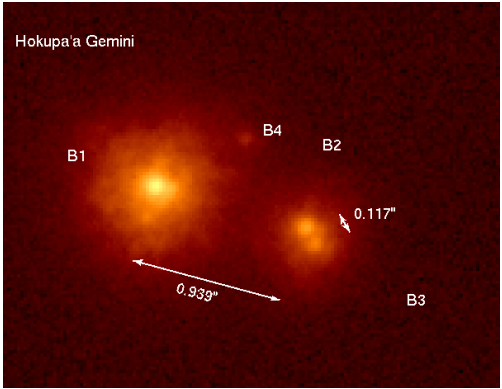}\caption{
The 8m Gemini/Hokupa'a images of the $\theta^{1}$ Ori B group in the $K^\prime$ band (09/19/01; from \cite{clo03c}). Resolution $0.085\arcsec$. Log scale. North is up and East is left. 
}
\label{fig4}
\end{figure}

\begin{figure}
\includegraphics[angle=0,width=0.4\columnwidth]{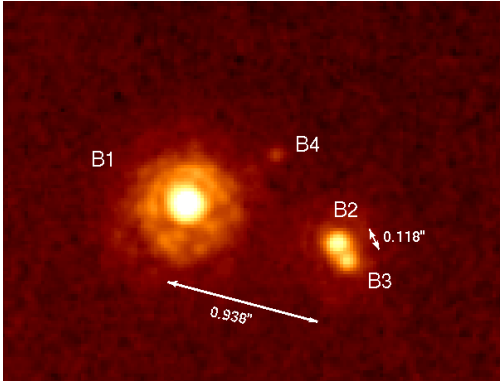}
\caption{
Detail of the $\theta^{1}$ Ori B group as imaged at $0.077\arcsec$ (Strehl $> 20\%$)
resolution (in the H band) with the MMT AO system (01/20/03) from \cite{clo03c}.}
\label{fig3b}
\end{figure}

\begin{figure}
\includegraphics[angle=0,width=0.6\columnwidth]{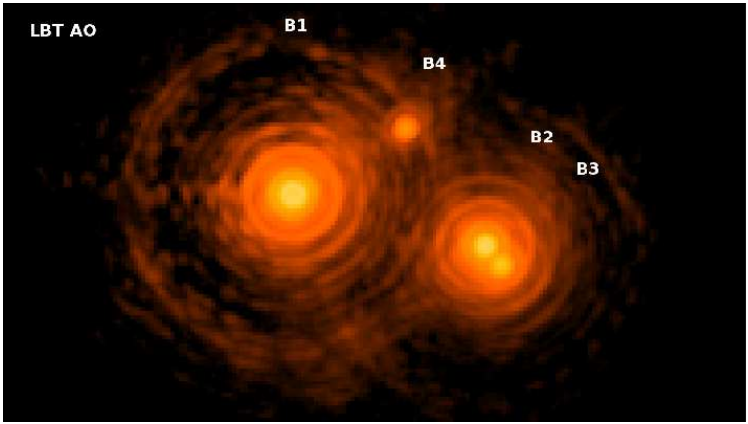}\caption{
The LBT AO $Br\gamma$ images of the $\theta^{1}$ Ori B group. Resolution $0.06\arcsec$. Logarithmic color scale. North is up and east is left. Strehl is $\sim 75\%$. 
}
\label{lbt_B}
\end{figure}

We can also compare our LBT data to older (somewhat less accurate) images of the Trapezium B stars from \cite{clo03c} who used AO images from Gemini and the 6.5m MMT and speckle images from the literature \citep{sch03}. Even though these individual observations are of lower quality and Strehl than the LBT ones (compare Figs. \ref{fig4} and \ref{fig3b} to that of the LBT in Fig. \ref{lbt_B}), the 15 years between these observations and those of the LBT can highlight even very small orbital motions of bound systems in the Trapezium. It also shows the very significant improvement in high Strehl AO now possible with Pyramid wavefront sensors and next generation adaptive secondary mirrors (ASMs).

\begin{figure}
\includegraphics[angle=0,width=0.6\columnwidth]{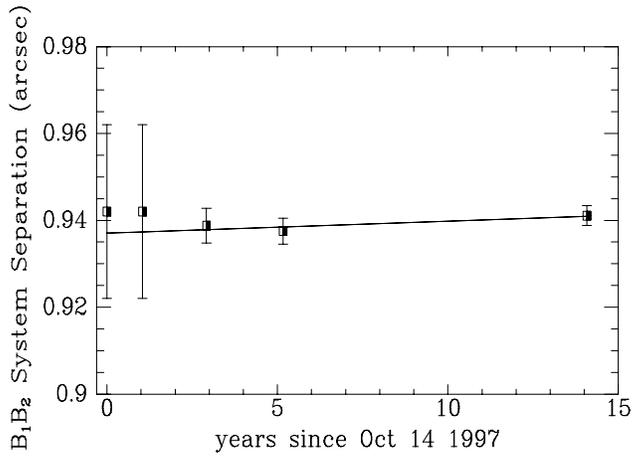}\caption{
The separation between $\theta^{1}$ Ori $B_1$ and $B_2$. Note how over
15 years of observation there has been little significant relative
proper motion observed (-0.27$\pm0.33 $ mas/yr; which is
a significant correlation only at 84\% level). If the group is
gravitationally bound the separation change could be this low. The first 2 data points are speckle
observations from the 6-m SAO telescope \citep{wei99}, the next point
is from Gemini/Hokupa'a observations \cite{clo03c} and the next data point is
from the MMT AO observations \cite{clo03c}, and the last from the LBT. }
\label{fig6}
\end{figure}

\begin{figure}
\includegraphics[angle=0,width=0.6\columnwidth]{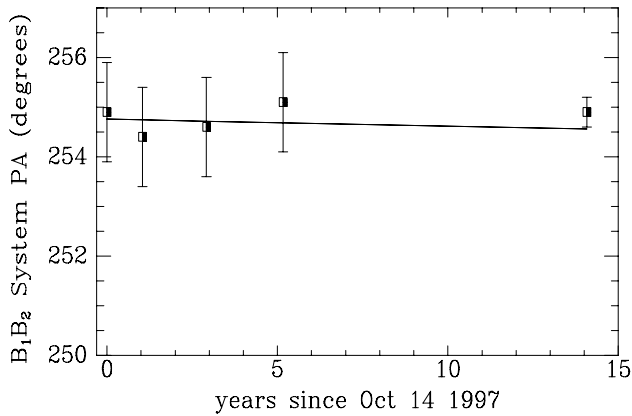}\caption{
The position angle between $\theta^{1}$ Ori $B_1$ and $B_2$. Note how
over 15 years of observation there has been no significant relative
PA motion observed (0.014$\pm0.048^\circ$/yr which is insignificantly different from zero velocity). The epochs of the data are the same as in Fig. \ref{fig6}.}
\label{fig7}
\end{figure}

\begin{figure}
\includegraphics[angle=0,width=0.6\columnwidth]{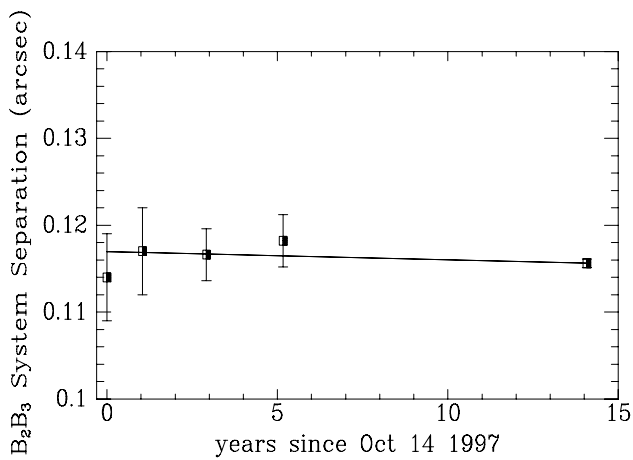}\caption{
The separation between $\theta^{1}$ Ori $B_2$ and $B_3$. Note the lack
of any significant relative motion ($-0.095\pm0.169$ mas/yr). The rms scatter from a constant value is only $0.16$ mas/yr. There appears to very little change in the separation of the $B_2B_3$ system. The epochs of the data are the same as in Fig. \ref{fig6}. }
\label{fig8}
\end{figure}

\begin{figure}
\includegraphics[angle=0,width=0.6\columnwidth]{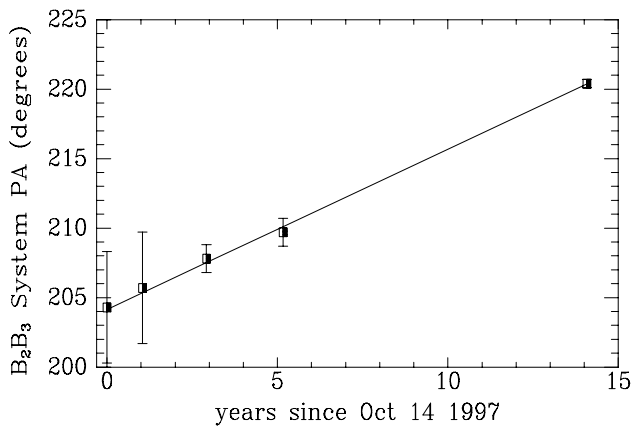}\caption{
The position angle of $\theta^{1}$ Ori $B_2$ and $B_3$. Here we
observe real orbital motion of $B_3$ moving
counter-clockwise (at 1.15$\pm0.07^\circ$/yr; correlation significant at the
99.9\% level) around $B_2$. This small amount of motion is consistant
with the $B_2B_3$ system being bound. The epochs of the data are the same as in Fig. \ref{fig6}. }
\label{fig9}
\end{figure}

\begin{figure}
\includegraphics[angle=0,width=0.5\columnwidth]{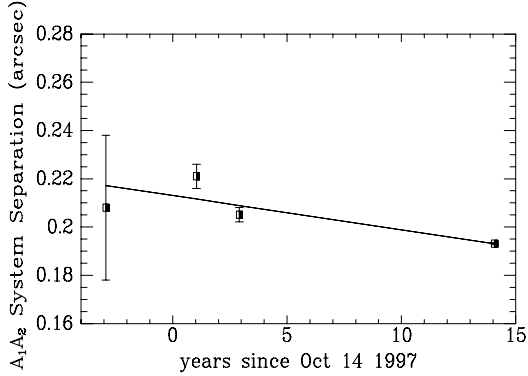}\caption{
The separation between $\theta^{1}$ Ori $A_1$ and $A_2$. There is a small negative change in the orbital separation
($-1.4\pm0.22$ mas/yr; correlation 94\%) as $A_2$ moves towards $A_1$. The first data point is from speckle observations
at the 3.5-m Calar Alto telescope \citep{pet98}, the next point is from
a speckle observation from the 6-m SAO telescope
\citep{wei99}, the next point is from the Gemini/Hokupa'a
observation of \cite{clo03c}, and the last point is from our LBT observations. }
\label{fig10}
\end{figure}

\begin{figure}
\includegraphics[angle=0,width=0.5\columnwidth]{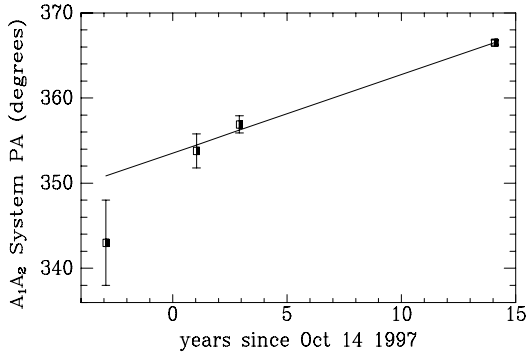}\caption{
The position angle of $\theta^{1}$ Ori $A_1$ and $A_2$. There is significant (linear correlation=99\%) change in the position angle as $A_2$ moves
counter clockwise (at 0.92$\pm0.07^\circ$/yr) around $A_1$. This
relatively large motion is consistent with the $A_1A_2$ system being
bound. The first data point is from speckle observations at the 3.5-m
Calar Alto telescope \citep{pet98}, the next point is from a speckle
observation from the 6-m SAO telescope \citep{wei99}, the next point
is from the Gemini/Hokupa'a observations of \cite{clo03c} and the last point is from our LBT observations. }
\label{fig11}
\end{figure}

\begin{figure}
\includegraphics[angle=0,width=0.6\columnwidth]{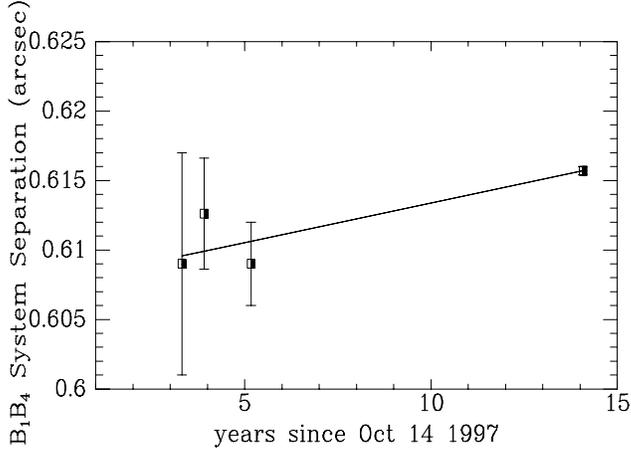}\caption{
The separation between $\theta^{1}$ Ori $B_1$ and $B_4$. Note how over
12 years of observation there been only now significant relative
proper motion observed (0.53$\pm0.24$ mas/yr; correlation 91\%). If the low mass star $B_4$
is gravitationally bound to the B group the $B_1B_4$ separation should
be roughly changing at a rate of this order.  The first data point is an speckle observation from the
6-m SAO telescope \citep{sch03}, the next is from the Gemini/Hokupa'a observation \cite{clo03c}
and the next data point is from the MMT AO observation \cite{clo03c} and the last is from the LBT. }
\label{figB4_sep}
\end{figure}

\begin{figure}
\includegraphics[angle=0,width=0.6\columnwidth]{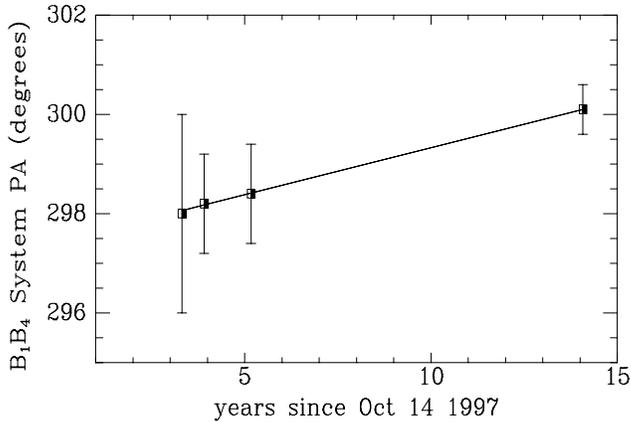}\caption{
The position angle between $\theta^{1}$ Ori $B_1$ and $B_4$. Note how
over 12 years of observation there has been only now a detectable significant relative
proper motion observed (0.18$\pm0.08^\circ$/yr; correlation 99.7\%). The sources of the data is the same as in Fig. \ref{figB4_sep}. }
\label{figB4_pa}
\end{figure}

A test to see how accurate our astrometry is over the last 15 years is to look at the separation and PA of B1 vs. B2. The scatter of the $\theta^{1}$ Ori $B_{1}B_{2}$ separation (which
should be very close to a constant since the $B_{1}B_{2}$ system has
an orbital period of $\ga 4000$ yr) will highlight systematic errors. The lack of any motion between $B_1$ and $B_2$ is also confirmed by \cite{sch03} and \cite{clo03c}. Our detailed LBT and past data on the B and A groups from the literature is summarized in Table
\ref{tbl-1}. Linear (weighted) fits to the
data in Table \ref{tbl-1} (Figures \ref{fig6} to \ref{figB4_pa})
yield the velocities shown in Table \ref{tbl-1}. The overall error in
the relative proper motions observed is now $\la 0.5$ mas/yr in
proper motion ($\la 1$ km/s).

\section{ANALYSIS \& DISCUSSION}

With these accuracies it is now possible to determine whether these
stars in the $\theta^{1}$ Ori $B$ group are bound together, or merely
chance projections in this very crowded region. We adopt the masses of each star from the \cite{sie97,ber96} tracks
fit by \cite{wei99} where we find masses of: $B_1\sim7M_{\sun} $;
$B_2\sim3M_{\sun} $; $B_3\sim2.5M_{\sun} $; $B_4\sim0.2M_{\sun} $;
$B_5\sim 7 M_{\sun}$; $A_1\sim20M_{\sun} $; $A_2\sim 4M_{\sun} $; and
$A_3\sim 2.6 M_{\sun}$. Based on these masses (which are similar to those adopted by \cite{sch03}) we can comment on
whether the observed motions are less than the escape velocities
expected for simple face-on circular orbits.

Our combination of high spatial resolution and high signal to noise
shows that there is no detectable motion in the $B_1B_2$ system over
the last 15 years (as we would expect if the true separation is
$\ga900$ AU)). But we have observed clear orbital
motion (at $4.9\pm0.3$ km/s) in the very tight $\theta^{1}$ Ori $B_{2}B_{3}$ system in PA
(see Figure \ref{fig9}). Also we see clear orbital motion (consistent with curvature) of $7.2\pm0.8$ km/s in the $\theta^{1}$ Ori $A_{1}A_{2}$
system (see Figs. \ref{fig10} and \ref{fig11}). We know this is likely orbital motion since both binaries have moved in an nearly circular arc of $\Delta PA\sim15^{\circ}$ over the last 15 years with almost no change in the separation between components, hence it appears (at least with the limited amount of observed orbital phase) that both of these binaries are consistent with roughly circular, close to face-on, orbits. 

%Since the $B_{2}B_{3}$ system has a projected separation of 52 AU and a system %mass of $\sim 5.5M_{\sun}$ the escape velocity is $\sim14$ km/s which is much g%reater than the $4.9\pm0.3$ km/s observed. 

%and 94 AU
%separations; respectively.

\subsection{Is the $\theta^1$ Ori $B_2B_3$ System Physical?}

The relative velocity in the $\theta^{1}$ Ori $B_{2}B_{3}$ system (in
the plane of the sky) is now more accurate by $\sim7\times$ compared to that of \cite{clo03c}. Our new velocity of $4.9\pm0.3$ km/s is consistent, but with much lower errors, with the $\sim 4.2\pm2.1$ km/s of \cite{clo03c} (this velocity is in the azimuthal direction; see Figure \ref{fig9}). This is a reasonable
$V_{tan}$ since an orbital velocity of $\sim 6.7$ km/s is expected
from a face-on circular orbit from a $\sim5.5 M_{\sun}$ binary system
like $\theta^{1}$ Ori $B_{2}B_{3}$ with a 51 AU projected
separation (implying an orbital period of order $\sim200$ yr). It is worth noting that this velocity is also greater than
the $\sim 3$ km/s \cite{hil98} dispersion velocity of the cluster.
Hence it is most likely that these two $K^\prime =7.6$ and $K^\prime
=8.6$ stars (separated by just 0.115$\arcsec$) are indeed in orbit around each other. Moreover, there are only 10 stars known to have $K^\prime < 8.6$ in the inner
$30\times 30\arcsec$ (see Figure \ref{fig2}), we
can estimate that the chances of finding two bright ($K^\prime < 8.6$)
stars within $0.115\arcsec$ is a small $<10^{-4}$ probability.

Our observed velocity of $1.15\pm0.07^{\circ}$/yr is consistent (in both direction and magnitude) with
the $1.4^{\circ}$/yr observed by \cite{sch03} and the $0.93\pm0.49^{\circ}$/yr of \cite{clo03c}. This suggests that the
AO and speckle datasets are both detecting real motion, but the addition of the LBT dataset has reduced the errors by $\sim7\times$. Moreover, since
this motion is primarily azimuthal strongly suggests an orbital arc of $B_3$
orbiting $B_2$.

\subsection{Is the $\theta^1$ Ori $A_1A_2$ System Physical?}

We observe $7.2\pm0.8$ km/s of relative motion in the
$\theta^{1}$ Ori $A_{1}A_{2}$ system (mainly in the azimuthal direction; see Figure \ref{fig11}). This is higher than the average
dispersion velocity of $\sim 3$ km/s yet well below an estimated
escape velocity of the $\sim 20 M_{\sun}$ $A_{1}A_{2}$ system
(projected separation of 94 AU). Hence it is highly likely that these
two $K^\prime =6.0$ and $K^\prime =7.6$ stars (separated by just
0.193$\arcsec$) are indeed in orbit around each other. In addition, there are only 8 stars known to have $K^\prime < 7.6$ in the inner
$30\times 30\arcsec$ (see Figure \ref{fig2}), we
can estimate that the chances of finding two bright ($K^\prime < 8.6$)
stars within $0.19\arcsec$ is a small $<4\times 10^{-4}$ probability.

Our observed velocity of $7.2\pm0.8$ km/s is much more accurate (and lower) than the $16.5\pm5.7$ km/s found by \cite{clo03c}. Our new value is consistent (in both direction and magnitude) with
the $\sim 10.3$ km/s observed by \cite{sch03}. This again suggests that the
AO and speckle datasets are both detecting real motion of $A_2$ orbiting $A_1$.

\subsection{Is the $\theta^1$ Ori B Group Stable?}

The pair $B_1B_5$ is moving at a low $\la 1$ km/s in the
plane of the sky w.r.t. to the pair $B_2B_3$ where the escape 
velocity $V_{\mathrm esc} \sim 6$\,km/s for this system. Hence these
pairs are very likely gravitationally bound together. However,
radial velocity measurements will be required to be absolutely
sure that these 2 pairs are truly bound together.

\subsubsection{Is the Orbit of $\theta^1$ Ori $B_4$ Stable?}

The situation is somewhat different for the faintest component of the
group, $B_4$. It has $K = 11.66$ mag which according to Hillenbrand \&
Carpenter (2000) suggests a mass of only $\sim 0.2\,{\mathrm
M}_\odot$. Since there are only 20 stars known to have $K < 11.66$ in
the inner $30 \time 30^{\prime\prime}$ (see Fig.\ 6), we can estimate
that the chances of finding a $K < 11.66$ star within
$0.6^{\prime\prime}$ of $B_1$ is a small $< 8 \times 10^{-3}$
probability. The two AO measurements of \cite{clo03c} (and the one speckle detection of \cite{sch03}) did not detect a
significant velocity of $B_4$ w.r.t. $B_1$: $2\pm11$ km/s.

However, our much better data and timeline has shed some light on the question of B4 orbiting B1. As is clear from Figures \ref{figB4_sep} \& \ref{figB4_pa}, the there appears to be a real velocity of $4.3\pm2.0$ km/s detected. This is greater than the random velocity of the cluster yet below the escape velocity of $\sim 6$ km/s, this points towards $B_4$ being gravitationally bound member of the $\theta^1$ Ori B group. This is the first time we can say that the lowest mass member of the B group is likely bound.

On the other hand, its very low mass and its projected location w.r.t. to the other four
groups members makes it highly unlikely that $B_4$ is on a long-term stable
orbit within the group. As we will discuss in the next section even the much more massive $B_3$ may not be stable in the long-term.

\subsubsection{Is the orbit of $B_3$ around $B_2$ and of $B_5$ around
$B_1$ stable in the long-term?}

$B_{1}B_{5}$, and $B_{2}B_{3}$ are two binaries with 
projected separations of 0.13 AU ($B_1B_5$) and 52 AU ($B_2B_3$); respectively. The two pairs are separated by a projected distance of 415 AU.
The distance $D_{B_1B_5} \sim 3 \times 10^{-4} \times D_{B_1B_5 B_2B_3}$ and thus
the $B_1B_5$ system is stable. Much more interesting is the case of $B_2B_3$. Their projected distance is not very small compared to their
projected distance (D) from the $B_1B_5$ pair:$ D_{B_2B_3} \sim 0.12 \times D_{B_1B_5 B_2B_3}$. Thus the stability of the $B_2B_3$ orbit needs a more
detailed analysis since it is possible that $B_3$ may be ejected in the future.

\cite{egg95} have given an empirical criterion for the
long-term stability of the orbits of hierarchical triple systems,
based on the results of their extensive model calculations
\citep{kis94,kis94a,egg95}. Their analytic stability criterion is good
to about $\pm 20\%$, and is meant to indicate stability for another
$10^2$ orbits. Given the uncertainties of the masses of the members of
the B group, this accuracy is sufficient for our present discussion.

The orbital period of the two binaries w.r.t. each other is 
$P_{(15)/(23)} \sim 1920\,$yrs, while the orbital period of $B_3$ w.r.t 
$B_2$ amounts to $P_{2/3} \sim 160\,$yrs. For the calculation of both 
periods, we have assumed the masses as given above, and circular orbits 
in the plane of the sky. This leads to a period ratio $X = 
P_{(15)/(23)}/P_{2/3} \sim 12$. Eggelton \& Kiseleva's stability 
criterion requires $X \ge  X_{\mathrm crit} = 10.08$ for the masses 
in the B group. This means that within the accuracy limits of our 
investigation, the binary $B_2B_3$ is just at the limit of 
stability. The stability criterion depends also on the orbits' 
eccentricities. In our case, already mild eccentricities of the order
of $e \sim 0.1$ (as can be expected to develop in hierarchical triple 
systems; see, e.g., Georgakarakos 2002), make the B group unstable.
While we cannot decide yet whether the pair $B_2B_3$ orbit each other
in a stable way, it is safe to say that that the ``triple'' $B_{1}B_{5}$, $B_2$,
and $B_3$ is not a simple, stable hierarchical triple system.

The $\theta^1$ Ori B system seems to be a good example of a highly
dynamic star formation "mini-cluster" which might in the future
eject the lowest-mass member(s) through dynamical decay
\citep{dur01}, and breaking up the gravitational binding of the widest
of the close binaries (the $B_{2}B_{3}$ system). The "ejection" of the lowest-mass member of a
formation "mini-cluster" could play a major role in the formation of
low mass stars and brown dwarfs \citep{rei01a,bat02,dur01,clo03a}. The
breaking up of binaries, of course, modifies the binary fraction of
main sequence stars considerably as well.

\section {FUTURE OBSERVATIONS}

Future observations are required to see, if indeed, these stars continue
to follow orbital arcs around each other proving that they are interacting
with one another. In addition, future observations of the $\theta^1$
Ori $B_4$ positions would help deduce if it is on a marginally stable orbit given its ``non-hierarchical'' location in the B group. 

Future observations should also try to determine the radial velocities
of these stars. Once radial velocities are known one can calculate
the full space velocities of these stars. Such observations will
require both very high spatial and spectral resolutions. This might be
possible with such instruments like the AO fed ARIES echelle
instrument at the MMT.

\acknowledgements

We thank the anonymous referee for their helpful review of this paper. The authors thank Piero Salinari for his insight, leadership and
persistence which made the development of the LBT adaptive secondaries
possible. We thank the whole LBT community for making this wonderful telescope possible. LMC is supported by an NSF AAG and NASA Origins of Solar Systems grants.

%% Tables should be submitted one per page, so put a \clearpage before

%% deluxetable environment provided by the AASTeX package or the LaTeX
%% table environment.  Use of deluxetable is preferred.
%%

%% Three table samples follow, two marked up in the deluxetable environment,
%% one marked up as a LaTeX table.

%% In this first example, note that the \tabletypesize{}
%% command has been used to reduce the font size of the table.
%% Note also that the \label command needs to be placed 

\clearpage
\begin{deluxetable}{llllllllll}
\tabletypesize{\scriptsize}
\tablecaption{Astrometry and Narrowband Photometry of the Trapezium Cluster, Oct 16/2011, LBT\label{tbl-0}}
\tablewidth{0pt}
\tablehead{
\colhead{RA\tablenotemark{a}} &
\colhead{DEC\tablenotemark{a}} &
\colhead{X} &
\colhead{Y} &
\colhead{$Br\gamma$} &
\colhead{Phot} &
\colhead{[FeII]} &
\colhead{Phot} &
\colhead{[FeII]-$Br\gamma$} &
\colhead{Comm.}\\
\colhead{J2000} &
\colhead{J2000} &
\colhead{pixel\tablenotemark{b}} &
\colhead{pixel\tablenotemark{b}} &
\colhead{(mag)} &
\colhead{Error} &
\colhead{(mag)} &
\colhead{Error} &
\colhead{Color} &
\colhead{}
}
\startdata
15.2934 & 23:23.1241 & 1973.024 & 1338.378 & 15.326 & 0.016 & 0 & 0 & na & no [FeII] image\\
15.3534 & 23:24.0418 & 1926.703 & 1290.952 & 16.102 & 0.012 & 0 & 0 & na &  no [FeII] image\\
15.5336 & 23:15.6412 & 1787.666 & 1725.089 & 16.035 & 0.015 & 17.715 & 0.024 & 1.68 & \\
15.5517 & 23:29.5216 & 1773.713 & 1007.755 & 16.859 & 0.015 & 0 & 0 & na & no [FeII] image \\
15.594 & 22:58.832 & 1741.035 & 2593.750 & 14.702 & 0.017 & 15.664 & 0.025 & 0.962 & \\
15.6306 & 22:56.385 & 1712.828 & 2720.215 & 10.676 & 0.012 & 11.945 & 0.009 & 1.269 & \\
15.7255 & 23:22.4347 & 1639.552 & 1374.004 & 11.743 & 0.007 & 12.970 & 0.032 & 1.227 & \\
15.7673 & 23:9.82764 & 1607.314 & 2025.533 & 9.3918 & 0.016 & 9.447 & 0.020 & 0.0552 & {\bf \bf{E1}} single\\
15.7879 & 23:26.5168 & 1591.355 & 1163.044 & 12.641 & 0.009 & 13.895 & 0.032 & 1.254 & $Br\gamma$ tail away from \bf{C1}\\
15.8018 & 23:11.8906 & 1580.662 & 1918.921 & 14.409 & 0.041 & 14.971 & 0.033 & 0.562 & \\
15.8202 & 23:14.2891 & 1566.428 & 1794.966 & 8.862 & 0.040 & 8.784 & 0.036 & -0.078 & {\bf A1} see table 2\\
15.8217 & 23:14.0972 & 1565.298 & 1804.883 & 10.322 & 0.037 & 10.392 & 0.033 & 0.07 & {\bf A2} see table 2\\
15.8337 & 23:22.4207 & 1556.059 & 1374.727 & 11.521 & 0.005 & 12.827 & 0.031 & 1.306 & \\
15.8408 & 23:25.5078 & 1550.599 & 1215.191 & 13.268 & 0.011 & 14.388 & 0.047 & 1.12 & $Br\gamma$ tail away from \bf{C1}\\
15.863 & 23:10.7606 & 1533.468 & 1977.318 & 14.411 & 0.023 & 15.485 & 0.041 & 1.074 & \\
15.8739 & 23:1.89992 & 1524.986 & 2435.234 & 12.637 & 0.014 & 13.553 & 0.031 & 0.916 & \\
15.9654 & 23:22.6589 & 1454.430 & 1362.420 & 16.199 & 0.017 & 17.120 & 0.025 & 0.921 & \\
16.0635 & 23:24.2937 & 1378.699 & 1277.933 & 15.907 & 0.065 & 17.586 & 0.082 & 1.679 & $Br\gamma$ tail away from \bf{C1}\\
16.064 & 23:7.05258 & 1378.302 & 2168.947 & 12.067 & 0.010 & 11.919 & 0.019 & -0.148 & {\bf B3} see table 2\\
16.069 & 23:6.96452 & 1374.429 & 2173.498 & 10.025 & 0.011 & 10.845 & 0.017 & 0.82 & {\bf B2} see table 2\\
16.0715 & 23:27.7444 & 1372.539 & 1099.602 & 15.928 & 0.027 & 17.152 & 0.062 & 1.224 & $Br\gamma$ tail away from \bf{C1}\\
16.0717 & 22:54.2677 & 1372.379 & 2829.663 & 14.591 & 0.015 & 18.480 & 0.049 & 3.889 & --$0.259\arcsec$ binary $\#$1B Table 3\\
16.0795 & 22:54.036 & 1366.341 & 2841.632 & 14.215 & 0.017 & 16.958 & 0.021 & 2.743 & --binary $\#$1A see Table 3\\
16.0928 & 23:23.0106 & 1356.092 & 1344.243 & 16.379 & 0.021 & 20.27 & 0.15 & 3.891 & very red Brown Dwarf?\\
16.0942 & 23:6.41047 & 1355.015 & 2202.131 & 13.552 & 0.024 & 13.825 & 0.029 & 0.273 & {\bf B4} see table 2\\
16.1006 & 23:14.1407 & 1350.043 & 1802.635 & 13.950 & 0.010 & 15.228 & 0.040 & 1.278 & --$0.187\arcsec$ bin. $\#$2A, Table 3\\
16.1014 & 23:14.2772 & 1349.480 & 1795.580 & 17.643 & 0.077 & 19.004 & 0.197 & 1.361 & -- bin. $\#$2B, Table 3\\ 
16.1299 & 23:6.71895 & 1327.477 & 2186.189 & 8.787 & 0.001 & 8.842 & 0.002 & 0.055 & {\bf B1} SB see table 2\\
16.1396 & 22:55.249 & 1319.930 & 2778.926 & 14.945 & 0.016 & 16.152 & 0.023 & 1.207 & \\
16.2263 & 23:19.0612 & 1253.022 & 1548.345 & 15.072 & 0.014 & 15.861 & 0.089 & 0.789 & \\
16.283 & 23:16.512 & 1209.290 & 1680.088 & 12.199 & 0.010 & 13.298 & 0.058 & 1.099 & astrometric zeropoint\tablenotemark{c}\\
16.3206 & 23:22.5317 & 1180.291 & 1368.992 & 15.185 & 0.029 & 16.261 & 0.029 & 1.076 & $2.115\arcsec$ proj. sep. to \bf{C1}\\
16.3241 & 23:25.2679 & 1177.537 & 1227.588 & 15.709 & 0.024 & 16.219 & 0.058 & 0.51 & \\
16.3997 & 23:11.2870 & 1119.16 & 1950.11 & 17.12 & 0.2 & 19.196 & 0.070 & 2.076 & Brown Dwarf? \\
16.4602 & 23:22.8832 & 1072.497 & 1350.827 & 8 & 1.0 & 8 & 1.0 & 0 & {\bf C1} binary (saturated)\\
16.4619 & 23:22.8443 & 1071.207 & 1352.838 & 7 & 1.0 & 7 & 1.0 & 0 & {\bf C2} $0.046\arcsec$ to \bf{C1}\\
16.6148 & 23:16.0836 & 953.204 & 1702.226 & 12.762 & 0.012 & 14.106 & 0.062 & 1.344 & \\
16.6529 & 23:28.8309 & 923.797 & 1043.453 & 14.928 & 0.011 & 16.050 & 0.058 & 1.122 & \\
16.7236 & 23:25.1688 & 869.219 & 1232.707 & 11.705 & 0.012 & 11.573 & 0.072 & -0.132 & \\
16.7469 & 23:16.3777 & 851.247 & 1687.030 & 11.721 & 0.009 & 13.529 & 0.061 & 1.808 & $Br\gamma$ tail away from \bf{C1}\\
16.7621 & 23:28.0209 & 839.571 & 1085.313 & 13.838 & 0.016 & 14.254 & 0.067 & 0.416 & \\
16.8258 & 23:25.9032 & 790.355 & 1194.754 & 16.951 & 0.045 & 18.094 & 0.075 & 1.143 & --$0.396\arcsec$ bin. $\#$3B, Table 3\\
16.8409 & 23:26.2297 & 778.753 & 1177.879 & 15.656 & 0.109 & 16.831 & 0.091 & 1.175 & --$Br\gamma$ Bowshock Bin $\#$3A\\
16.8633 & 23:7.03118 & 761.435 & 2170.053 & 15.402 & 0.008 & 16.107 & 0.066 & 0.705 & \\
17.0595 & 23:33.9787 & 610.024 & 777.417 & 11.020 & 0.030 & 11.264 & 0.035 & 0.244 & \\
17.1675 & 23:17.0013 & 526.70 & 1654.80 & 0.0 & 0.0 & 15.465 & 0.107 & na &{\bf D2} $1.401\arcsec$ proj. sep. to \bf{D1}\\
17.2558 & 23:16.5298 & 458.479 & 1679.17 & 0.0 & 0.0 & 8.658 & 0.1 & na &{\bf D1} no $Br\gamma$ image\\
\enddata
\tablenotetext{a}{to calculate full RA simply prefix 5:35 to column 1 ($True_{RA}="5:35:col1"$) and for full DEC simply prefix -5: to column 2 ($True_{DEC}="-5:col2$''). So for example we find $\theta^{1}$ Ori C1 is at RA=5:35:16.4602 and DEC=-5:23:22.8832 (J2000).}
\tablenotetext{b}{these pixels are in the $Br\gamma$ filter with a platescale of $0.019350\pm0.000047\arcsec /pix$. Increasing Y is due North, and increasing X is in the due West direction}
 \tablenotetext{c}{we used this single star (called LV3 or 163-317) as the astrometric zeropoint as it has a good position the the Hubble ACS data of Ricci et al. and is imaged in all of of our AO data. It has a location of 5:35:16.460 -5:23:22.812 J2000.}  
\end{deluxetable}

\clearpage
\begin{deluxetable}{lllllllll}
\tabletypesize{\scriptsize}
\tablecaption{High Resolution Observations of the $\theta^{1}$ Ori B \& A groups\label{tbl-1}}
\tablewidth{0pt}
\tablehead{
\colhead{System} &
\colhead{$\Delta H$} &
\colhead{$\Delta K^{\prime}$} &
\colhead{Separation} &
\colhead{Sep. Vel.} &
\colhead{PA} &
\colhead{PA Vel.} &
\colhead{Telescope} &
\colhead{epoch}\\
\colhead{name} &
\colhead{(mag)} &
\colhead{(mag)} &
\colhead{($\arcsec $)} &
\colhead{(Sep. mas/yr)} &
\colhead{($^{\circ}$)} &
\colhead{($^{\circ}$/yr)} &
\colhead{} &
\colhead{(m/d/y)}\\
}
\startdata

$B_{1}B_{2}$&$2.30\pm0.15$&&$0.942\pm0.020\arcsec$&&$254.9\pm1.0$&&SAO\tablenotemark{a}&10/14/97\\
&&$1.31\pm0.10$\tablenotemark{b}&$0.942\pm0.020\arcsec$&&$254.4\pm1.0$&&SAO\tablenotemark{a}&11/03/98\\
&&$2.07\pm0.05$&$0.9388\pm0.0040\arcsec$&&$255.1\pm1.0$&&GEMINI&09/19/01\\
&$2.24\pm0.05$&&$0.9375\pm0.0030\arcsec$&&$255.1\pm1.0$&&MMT&01/20/03\\
%&&&$<0.9400\pm0.0019>$&&$<254.9\pm0.3>$&&&\\
&&&$0.9411\pm0.0023\arcsec$&&$254.5\pm0.3$&&LBT&10/16/11\\
&&&&&&&&\\
&&& before LBT= &-0.6$\pm1.9$&&0.07$\pm0.25$&&\\
&&& with LBT=&-0.27$\pm0.33$&&-0.013$\pm0.048$&&\\
&&&corr.=& 84\%; no vel. & corr.=& 45\%; no vel. &&\\
&&&& detected && detected &&\\
\hline
&&&&&&&&\\
%$B_{1}B_{3}$&$3.1\pm0.7$&&$1.018\pm0.020\arcsec$&&$250.0\pm1.0$&&SAO\tablenotemark{a}&10/14/97\\
%&&$2.5\pm0.40$\tablenotemark{b}&$1.023\pm0.020\arcsec$&&$249.5\pm1.0$&&SAO\tablenotemark{a}&11/03/98\\
%&&$3.11\pm0.05$&$1.022\pm0.0040\arcsec$&&$249.8\pm1.0$&&GEMINI&09/19/01\\
%&$3.08\pm0.05$&&$1.024\pm0.0030\arcsec$&&$250.4\pm1.0$&&MMT&01/19/03\\
%&&&$<1.021\pm0.0023>$&&$<259.9\pm0.3>$&&&\\
%&&&&-0.0006$\pm0.0019\arcsec$/yr&&0.07$\pm0.25^\circ$/yr&&\\
%&&&&&&&&\\
$B_{2}B_{3}$&$1.00\pm0.11$&&$0.114\pm0.05\arcsec$&&$204.3\pm4.0$&&SAO\tablenotemark{a}&10/14/97\\
&&$1.24\pm0.20$&$0.117\pm0.005\arcsec$&&$205.7\pm4.0$&&SAO\tablenotemark{a}&11/03/98\\
&&$1.04\pm0.05$&$0.1166\pm0.0040\arcsec$&&$207.8\pm1.0$&&GEMINI&09/19/01\\
&$0.85\pm0.05$&&$0.1182\pm0.0030\arcsec$&&$209.7\pm1.0$&&MMT&01/20/03\\
&&&$0.1156\pm0.0005\arcsec$&&$220.4\pm0.3$&&LBT&10/16/11\\
&&&&&&&&\\
&&&before LBT= &$0.6\pm1.0$ &&0.93$\pm0.49$ &&\\
&&&with LBT=&-0.10$\pm0.16$&&1.15$\pm0.07$&&\\
&&&corr.=&56\%; no vel.&&{\bf 4.9$\pm$0.3 km/s}&&\\
&&&& detected && {\bf corr.=99.9\%} &&\\
\hline
&&&&&&&&\\
$B_{1}B_{4}$&&$5.05\pm0.8$&$0.609\pm0.008\arcsec$&&$298.0\pm2.0$&&SAO\tablenotemark{c}&02/07/01\\
&&$5.01\pm0.10$&$0.6126\pm0.0040\arcsec$&&$298.2\pm1.0$&&GEMINI&09/19/01\\
&$4.98\pm0.10$&&$0.6090\pm0.0050\arcsec$&&$298.4\pm1.0$&&MMT&01/20/03\\
&&&$0.6157\pm0.003\arcsec$&&$300.1\pm0.5$&&LBT&10/16/11\\
&&&&&&&&\\
&&&before LBT=&$-1.1\pm1.9$ &&0.1$\pm0.5$&&\\
&&&with LBT=&0.53$\pm0.25$&&0.18$\pm0.08$&&\\
&&&&{\bf 1.1$\pm$0.5 km/s}&&{\bf 4.13$\pm$1.8 km/s}&&\\
&&&&{\bf corr.=91.1\%} && {\bf corr.=99.7\%} &&\\
\hline
&&&&&&&&\\
%$B_{2}B_{4}$&&$2.94\pm0.05$&$0.6512\pm0.0040\arcsec$&&$34.1\pm1.0$&&GEMINI&09/19/01\\
%&$2.74\pm0.10$&&$0.6475\pm0.0050\arcsec$&&$34.9\pm1.0$&&MMT&01/20/03\\
%&&&&-0.0037$\pm0.0064\arcsec$/yr&&0.8$\pm1.4^\circ$/yr&&\\ 
%\hline
%&&&&&&&&\\
$A_{1}A_{2}$&$1.51\pm0.15$&$1.38\pm0.10$&$0.208\pm0.030\arcsec$&&$343.5\pm5.0$&&Calar Alto\tablenotemark{d}&11/15/94\\
&&$1.51\pm0.05$&$0.2215\pm0.005\arcsec$&&$353.8\pm2.0$&&SAO\tablenotemark{a}&11/03/98\\
&&$1.62\pm0.05$&$0.2051\pm0.0030\arcsec$&&$356.9\pm1.0$&&GEMINI&09/19/01\\
&&&$0.1931\pm0.0005\arcsec$&&$366.5\pm0.3$&&LBT&10/16/11\\ 
&&&&&&&&\\
&&&before LBT=&$-6.4\pm2.7$ &&2.13$\pm0.73$ &&\\
&&&with LBT=&-1.4$\pm0.2$&&0.92$\pm0.07$&&\\
&&&&{\bf 2.9$\pm$0.4 km/s}&&{\bf 6.6$\pm$0.5 km/s}&&\\
&&&&{\bf corr.= 94.1\%}&&{\bf corr.= 98.9\%}&&\\
\enddata
\tablenotetext{a}{speckle observations of \cite{wei99}.}
\tablenotetext{b}{these low $\Delta K$ values are possibly due to $\theta^{1}$ Ori $B_{1}$ being in eclipse during the 11/03/98 observations of \cite{wei99}. }
\tablenotetext{c}{speckle observations of \cite{sch03}.}
\tablenotetext{d}{speckle observations of \cite{pet98}.}
\end{deluxetable}

\clearpage
\begin{deluxetable}{llllllllll}
\tabletypesize{\scriptsize}
\tablecaption{Tight Binaries in the LBT Trapezium Field (10/16/11)\label{tbl-2}}
\tablewidth{0pt}
\tablehead{
\colhead{Bin.} &
\colhead{RA\tablenotemark{a}} &
\colhead{DEC\tablenotemark{a}} &
\colhead{X} &
\colhead{Y} &
\colhead{$\Delta Br\gamma$} &
\colhead{$\Delta [FeII]$} &
\colhead{Sep.} &
\colhead{PA} &
\colhead{Comment} \\
\colhead{$\#$} &
\colhead{J2000} &
\colhead{J2000} &
\colhead{pixel\tablenotemark{b}} &
\colhead{pixel\tablenotemark{b}} &
\colhead{(mag)} &
\colhead{(mag)} &
\colhead{(mas)} &
\colhead{($^{\circ}$)} &
\colhead{}
}
\startdata
1A & 16.0795 & 22:54.036 & 1366.341 & 2841.632 & $0.376\pm0.024$ &$1.522\pm0.053$&$259\pm2$&$206.77\pm0.30$&B has more\\
1B & 16.0717 & 22:54.2677 & 1372.379 & 2829.663 &  &  & & & extinction than A\\
   &         &            &          &          &  &  & & & A\&B both very red \\
\hline
2A &16.1006 & 23:14.1407 & 1350.043 & 1802.635 & $3.693\pm0.080$ &$3.776\pm0.198$&$136\pm3$&$175.44\pm0.30$&B has strong\\ 
2B &16.1014 & 23:14.2772 & 1349.480 & 1795.580 & & & & &  gray extinction?\\
\hline
3A &16.8409 & 23:26.2297 & 778.753 & 1177.879 & $1.295\pm0.064$ &$1.263\pm0.118$&$396\pm3$&$325.49\pm0.30$&$Br\gamma$ Bowshock\\
3B &16.8258 & 23:25.9032 & 790.355 & 1194.754 & &  & & &between binary?\\
\enddata
\tablenotetext{a}{The RA and DEC are the same format as in Table \ref{tbl-0}}
\tablenotetext{b}{The X and Y location are same format as in Table \ref{tbl-0}}
\end{deluxetable}

%\clearpage

%\begin{figure}
% \includegraphics[angle=0,width=\columnwidth]{f0.eps}
%\caption{
%An exploded view of the new technology deformable secondary
%mirror. The active surface of the secondary is a very thin aspherical
%shell (1.8 mm thick, 65 cm in diameter) which is deformed by 336
%magnetic actuators. Effectively the shell is ``floated'' by a magnetic
%field with a $\sim$50 $\mu m$ air gap between its spherical
%backsurface and the identically polished spherical reference body
%(also made of ULE glass). The gap thickness is controlled to $\pm4$ nm
%by 336 voice coil actuators under servo control of the capacitive
%sensors sampling at 20 KHz. This secondary mirror (the world's first)
%is a joint project of University of Arizona and the Italian National
%Institute of Astrophysics - Arcetri Observatory.  }
%\label{fig0}
%\end{figure}

%\clearpage

%\begin{figure}
% \includegraphics[angle=0,width=\columnwidth]{f1_small.eps}
%\caption{
%The new technology deformable secondary mirror being installed at the 6.5 meter MMT telescope at Mt. Hopkins, Arizona. (shown
%from left to right: Michael Lloyd-Hart, Francois Wildi, \& Laird Close of Steward Observatory) }
%\label{fig1}
%\end{figure}

\clearpage

\clearpage

\clearpage

\clearpage

\clearpage

\clearpage

\clearpage

\clearpage

\clearpage

%% The following command ends yourmanuscript. LaTeX will ignore any text 
\end{document}